\documentclass[aps,prd,preprint,floatfix,nofootinbib,superscriptaddress]{revtex4}

\usepackage{graphicx,subfigure,epsfig,epsf}

\def\poutsla{/\!\!\!p_{out}}
\def\pinsla{/\!\!\!p_{in}}
\def\ponesla{/\!\!\!p_{1}}
\def\ptwosla{/\!\!\!p_{2}}

\def\pthreesla{/\!\!\!p_{3}}
\def\pfoursla{/\!\!\!p_{4}}

\newcommand{\beq}{\begin{equation}}
\newcommand{\eeq}{\end{equation}}
\newcommand{\bea}{\begin{eqnarray}}
\newcommand{\eea}{\end{eqnarray}}

\newcommand{\Lda}{\Lambda}
\newcommand{\g}{\gamma}
\def\Re{{\cal R \mskip-4mu \lower.1ex \hbox{\it e}\,}}
\def\Im{{\cal I \mskip-5mu \lower.1ex \hbox{\it m}\,}}

\def\etal{{\it et al.}}

\def\tev{\,{\ifmmode\mathrm {TeV}\else TeV\fi}}
\def\gev{\,{\ifmmode\mathrm {GeV}\else GeV\fi}}
\def\mev{\,{\ifmmode\mathrm {MeV}\else MeV\fi}}
\def\to{\rightarrow}

\begin{document}


\def\issue(#1,#2,#3){{\bf #1,} #2 (#3)} 
\def\APP(#1,#2,#3){Acta Phys.\ Pol.\ B \ \issue(#1,#2,#3)}
\def\ARNPS(#1,#2,#3){Ann.\ Rev.\ Nucl.\ Part.\ Sci.\ \issue(#1,#2,#3)}
\def\CPC(#1,#2,#3){Comp.\ Phys.\ Comm.\ \issue(#1,#2,#3)}
\def\CIP(#1,#2,#3){Comput.\ Phys.\ \issue(#1,#2,#3)}
\def\EPJC(#1,#2,#3){Eur.\ Phys.\ J.\ C\ \issue(#1,#2,#3)}
\def\EPJD(#1,#2,#3){Eur.\ Phys.\ J. Direct\ C\ \issue(#1,#2,#3)}
\def\IEEETNS(#1,#2,#3){IEEE Trans.\ Nucl.\ Sci.\ \issue(#1,#2,#3)}
\def\IJMP(#1,#2,#3){Int.\ J.\ Mod.\ Phys. \issue(#1,#2,#3)}
\def\JHEP(#1,#2,#3){J.\ High Energy Physics \issue(#1,#2,#3)}
\def\JPG(#1,#2,#3){J.\ Phys.\ G \issue(#1,#2,#3)}
\def\MPL(#1,#2,#3){Mod.\ Phys.\ Lett.\ \issue(#1,#2,#3)}
\def\NP(#1,#2,#3){Nucl.\ Phys.\ \issue(#1,#2,#3)}
\def\NIM(#1,#2,#3){Nucl.\ Instrum.\ Meth.\ \issue(#1,#2,#3)}
\def\PL(#1,#2,#3){Phys.\ Lett.\ \issue(#1,#2,#3)}
\def\PRD(#1,#2,#3){Phys.\ Rev.\ D \issue(#1,#2,#3)}
\def\PRL(#1,#2,#3){Phys.\ Rev.\ Lett.\ \issue(#1,#2,#3)}
\def\PTP(#1,#2,#3){Progs.\ Theo.\ Phys. \ \issue(#1,#2,#3)}
\def\RMP(#1,#2,#3){Rev.\ Mod.\ Phys.\ \issue(#1,#2,#3)}
\def\SJNP(#1,#2,#3){Sov.\ J. Nucl.\ Phys.\ \issue(#1,#2,#3)}

\begin{flushright}
BITSGoa-2007/10/002\\
\end{flushright}


\bibliographystyle{revtex}

\title{M\"{o}ller and Bhabha scattering in the  noncommutative \\
standard model } 




\author{P.K.Das}
\affiliation{Birla Institute of Technology and Science-Pilani, Goa campus, NH-17B, Zuarinagar, Goa-403726, India}

\author{N.G.Deshpande}
\affiliation{Institute for Theoretical Science, University of Oregon,
Eugene, OR 97403}

\author{G.Rajasekaran}
\affiliation{The Institute of Mathematical Sciences, C.I.T Campus, Taramani, Chennai 600113, India}


\date{\today}

\begin{abstract}
We study the M\"{o}ller and Bhabha scattering in the noncommutative extension of the standard model(SM) using the Seiberg-Witten maps of this to first order of the noncommutative parameter $\theta_{\mu \nu}$. We look at the angular distribution $d\sigma/d\Omega$ to explore the noncommutativity of space-time at around $\Lambda_{NC} \sim$ TeV and find that the distribution deviates significantly from the one obtained from the commutative version of the standard model. 
\end{abstract}


\maketitle


\section{Introduction}
Interest in the noncommutative field theory is old and it arose 
from the pioneering work by Snyder \cite{Snyder47} and has been revived
recently due to developments connected to string theories in which
the noncommutativity of space-time is an important characteristic
of D-brane dynamics at low energy limit\cite{Connes98,Douglas98,SW99,Schomerus}. 
Although Douglas \etal \cite{Douglas98} in their pioneering work have shown that noncommutative field theory is a well-defined quantum field theory, the question that remains whether string theory prediction and the noncommutative effect can be seen at the energy scale attainable in present or near future experiments instead of the $4$-$d$ Planck scale $M_{pl}$. A notable work by Witten \etal \cite{Witten96} suggests that one can see some stringy effects by lowering down the threshold value of commutativity to \tev, a scale which is not so far from present or future collider scale. 

~ What is space-time noncommutavity? It means space and time no longer commute with each other. Now writing the space-time coordinates as operators (same as the position and momentum operator in quantum mechanics) we find 
\beq [\hat{X}_\mu,\hat{X}
_\nu]=i\theta_{\mu\nu}=\frac{1}{\Lda_{NC}^2}i c_{\mu\nu}
\label{NCST}
\eeq
 where $c_{\mu \nu}$ are antisymmetric constant parameters 
and $\Lda_{NC}$ is the scale at which space-time is no longer commutative. To study an ordinary field theory in such a noncommutative space-time one replaces all ordinary products among the field variables with Moyal-Weyl(MW) 
\cite{Douglas} $\star$ products defined by
\begin{equation}
(f\star
g)(x)=exp\left(\frac{1}{2}\theta_{\mu\nu}\partial_{x^\mu}\partial_{y^\nu}\right)f(x)g(y)|_{y=x}.
\label{StarP}
\end{equation}
Using this we can get the NCQED Lagrangian as
\begin{equation} \label{ncQED}
{\cal L}=\frac{1}{2}i(\bar{\psi}\star \gamma^\mu D_\mu\psi
-(D_\mu\bar{\psi})\star \gamma^\mu \psi)- m\bar{\psi}\star
\psi-\frac{1}{4}F_{\mu\nu}\star F^{\mu\nu} \label{NCL},
\end{equation}
which are invariant under the following transformations 
\bea
\psi(x,\theta) \to \psi'(x,\theta) &=& U \star \psi(x,\theta), \\
A_{\mu}(x,\theta) \to A_{\mu}'(x,\theta) &=& U \star A_{\mu}(x,\theta) \star U^{-1} + \frac{i}{e} U \star \partial_\mu U^{-1},
\eea
where $U = (e^{i \Lambda})_\star$. In the NCQED lagrangian (Eq.\ref{ncQED})
$D_\mu\psi=\partial_\mu\psi-ieA_\mu\star\psi$,$~~(D_\mu\bar{\psi})=\partial_\mu\bar{\psi}+ie\bar{\psi}\star
A_\mu$, $~~ F_{\mu\nu}=\partial_{\mu} A_{\nu}-\partial_{\nu}
A_{\mu}-ie(A_{\mu}\star A_{\nu}-A_{\nu}\star A_{\mu})$. 

 The alternative one is the Seiberg-Witten(SW)\cite{SW99,Douglas98,Connes98,Jurco} approach in which both the gauge parameter $\Lambda$ and the gauge field $A^\mu$
are expanded as 
\bea \label{swps}
\Lambda_\alpha (x,\theta) &=& \alpha(x) + \theta^{\mu\nu} \Lambda^{(1)}_{\mu\nu}(x;\alpha) + \theta^{\mu\nu} \theta^{\eta\sigma} \Lambda^{(2)}_{\mu\nu\eta\sigma}(x;\alpha) + \cdot \cdot \cdot \\
A_\rho (x,\theta) &=& A_\rho(x) + \theta^{\mu\nu} A^{(1)}_{\mu\nu\rho}(x) + \theta^{\mu\nu} \theta^{\eta\sigma} A^{(2)}_{\mu\nu\eta\sigma\rho}(x) + \cdot \cdot \cdot
\eea
and when the field theory is expanded in terms of this power series (\ref{swps}) one end up with an infinite tower of higher dimensional operators which renders the theory nonrenormalizable. However, the advantage is that this construction can be applied to any gauge theory with arbitrary matter representation. In the MW 
approach the group closure property was found to hold only for 
$U(N)$ gauge theories with matter content in the fundamental or adjoint representations. 
~Using the SW-map, Calmet \etal \cite{Calmet} first constructed a model with noncommutative gauge invariance which was close to the usual Standard Model and is known as the {\it minimal} NCSM(mNCSM) and 
they listed several Feynman rules. Many phenomenological studies 
\cite{Hewett01} have been made to unravel several interesting features of this mNCSM. Hewett \etal explored several processes e.g. 
$e^+ e^- \to e^+ e^-$ (Bhabha), $e^- e^- \to e^- e^-$ (M\"{o}ller), 
$e^- \g \to e^- \g$, $e^+ e^- \to \g \g$ (pair annihilation), $\g \g \to e^+ e^-$ and $\g \g \to \g \g$ in context of NCQED. They found that the differential cross-section for Bhabha scattering(a s-channel process) is dependent on the space-time NC parameters $\theta_{0i},(i=1,2,3)$, whereas in M\"{o}ller scattering(a t and u-channel dominated process) the sensitivities are on $\theta_{12}$ and $\theta_{13}$, if the beam is in the 1-direction. However, their analyses were only in the context of NCQED, not in full mNCSM, i.e. they didn't consider the impact of the  neutral $Z$ boson exchange in Bhabha and M\"{o}ller scattering.  Here we consider the impact of both $Z$ and photon exchange in our analysis of the above two processes in the NC framework and will see the modification in the angular distribution of the differential cross-section which arise both from the polar($\theta^*$) and the azumuthal angular($\phi$) dependences. Note that in the earlier analyses only the $\phi$ distribution of the cross-section was studied, whereas in this work we study also the polar distribution. Since we have the $Z$ mediated diagrams, the interference term between the photon and $Z$ mediated Feynman diagrams, do have some impact on such distribution, which in other words is nothing but a measure of direct CP-asymmetry. Now in a generic NCQED the triple photon arises to order ${\mathcal{\theta}}$, which however is absent in this minimal mNCSM. Another formulation of the NCSM came in forefront through the pioneering work by  Melic \etal \cite{Melic:2005ep}
where such a triple neutral gauge boson coupling \cite{Trampetic} appears naturally in the gauge sector. We will call this the non-minimal version of NCSM or simply NCSM. In the present work we will confine ourselves within this non-minimal version of the NCSM and use the Feynman rules for interactions given in Melic \etal \cite{Melic:2005ep}.
 

\section{M\"{o}ller scattering in the NCSM}
The M\"{o}ller scattering $e^- (p_1) e^- (p_2) \rightarrow e^- (p_3) e^- (p_4)$ in the NCSM  proceeds via the $t$ and $u$ channel 
exchange of $\gamma$ and $Z$ boson just like the usual SM(which is a commutative one and is being described as CSM). The corresponding Feynman diagrams are shown in Fig. 1 
\begin{figure}[htbp]
\vspace{5pt}
\centerline{\hspace{-3.3mm}
{\epsfxsize=12cm\epsfbox{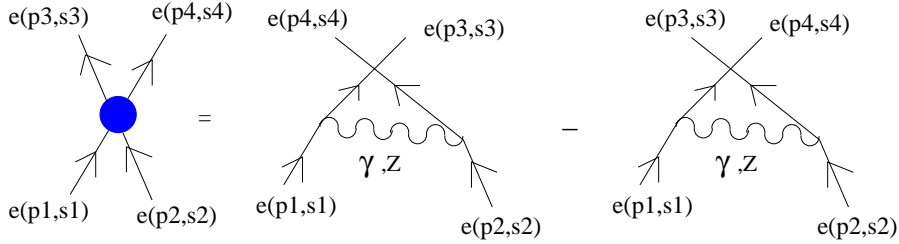}}}
\hspace{3.0cm}
\caption{Feynman diagrams for the M\"{o}ller scattering.}
\protect\label{fig1}
\end{figure}

\noindent The scattering amplitude to order 
$\theta$, for the photon mediated diagrams can be written as 
\bea
{\mathcal{A}}_{M}^\gamma = e^2 \left[(1 - \frac{i}{2} p_4 \theta p_2 + \frac{i}{2} p_3 \theta p_1) {\overline u}(p_3,s_3) \gamma^\mu u(p_1,s_1)  {\overline u}(p_4,s_4) \gamma_\mu u(p_2,s_2) \left(\frac{i}{t}\right) \right] \nonumber \\ 
- e^2 \left[ (1 - \frac{i}{2} p_3 \theta p_2 + \frac{i}{2} p_4 \theta p_1) {\overline u}(p_4,s_4) \gamma^\mu u(p_1,s_1)  {\overline u}(p_3,s_3) \gamma_\mu u(p_2,s_2) \left(\frac{i}{u}\right)\right] \nonumber \\
= {\mathcal{A}}_{1M}^\gamma - {\mathcal{A}}_{2M}^\gamma,
\eea
and for the $Z$ boson mediated diagrams as 
\bea
{\mathcal{A}}_{M}^Z = \frac{e^2}{x_W^2} \left[(1 - \frac{i}{2} p_4 \theta p_2 + \frac{i}{2} p_3 \theta p_1) {\overline u}(p_3,s_3) \gamma^\mu \Gamma_A^- u(p_1,s_1)  {\overline u}(p_4,s_4) \gamma_\mu \Gamma_A^- u(p_2,s_2) 
\left(\frac{i}{t - m_Z^2}\right) \right] \nonumber \\
- \frac{e^2}{x_W^2} \left[ (1 - \frac{i}{2} p_3 \theta p_2 + \frac{i}{2} p_4 \theta p_1) {\overline u}(p_4,s_4) \gamma^\mu \Gamma_A^- u(p_1,s_1)  {\overline u}(p_3,s_3) \gamma_\mu \Gamma_A^- u(p_2,s_2) \left(\frac{i}{u - m_Z^2}\right)\right] \nonumber \\
= {\mathcal{A}}_{1M}^Z - {\mathcal{A}}_{2M}^Z,
\eea
where $p_1 \theta p_2  = p_1^\mu \theta_{\mu \nu} p_2^\nu$, $p_1 + p_2 = p_3 + p_4$, $t=(p_1 - p_3)^2$, $u=(p_1 - p_4)^2$,
and $s=(p_1 + p_2)^2$. Also $x_W =sin 2\theta_W$($\theta_W$,the Weinberg angle), $\Gamma_A^\pm = c_V^e \pm c_A^e \gamma_5$ with $c_V^e = T_3^e - 2 Q_e sin^2 \theta_W$ and $c_A^e = T_3^e$. 

The spin-averaged and summed squared-amplitude is given by
\bea \label{eqn:MollerAmp}
{\overline {|{\mathcal{A}}_{M}|^2}} = {\overline {|{\mathcal{A}}_{1M}^\gamma|^2}} + {\overline {|{\mathcal{A}}_{2M}^\gamma|^2}}+ {\overline {|{\mathcal{A}}_{1M}^Z|^2}} + {\overline {|{\mathcal{A}}_{2M}^Z|^2}} - 2 {\overline {Re({\mathcal{A}}_{1M}^\gamma {\mathcal{A}}_{2M}^{\gamma *})}} - 2 {\overline {Re({\mathcal{A}}_{1M}^Z {\mathcal{A}}_{2M}^{Z *})}} + 2 {\overline {Re({\mathcal{A}}_{1M}^\gamma {\mathcal{A}}_{1M}^{Z *})}} 
\nonumber \\
 - 2 {\overline {Re({\mathcal{A}}_{1M}^\gamma {\mathcal{A}}_{2M}^{Z *})}} - 2 {\overline {Re({\mathcal{A}}_{2M}^\gamma {\mathcal{A}}_{1M}^{Z *})}} + 2 {\overline {Re({\mathcal{A}}_{2M}^\gamma {\mathcal{A}}_{2M}^{Z *})}} = \frac{1}{4} \sum_{spins} |{\mathcal{A}}_M^\gamma + {\mathcal{A}}_M^Z |^2.  \nonumber \\
\eea
Several terms in the squared-amplitude are given in Appendix B.

\section{Bhabha scattering in the NC standard model}
Next the Bhabha scattering $e^- (p_1) e^+ (p_2) \rightarrow e^- (p_3) e^+ (p_4)$. As in the usual SM, the Bhabha scattering in the NCSM 
proceeds via the $s$ and $t$ channel exchange of  
$\gamma$ and $Z$ bosons. The respective Feynman diagrams are shown in Fig. 2 
\begin{figure}[htbp]
\vspace{5pt}
\centerline{\hspace{-3.3mm}
{\epsfxsize=12cm\epsfbox{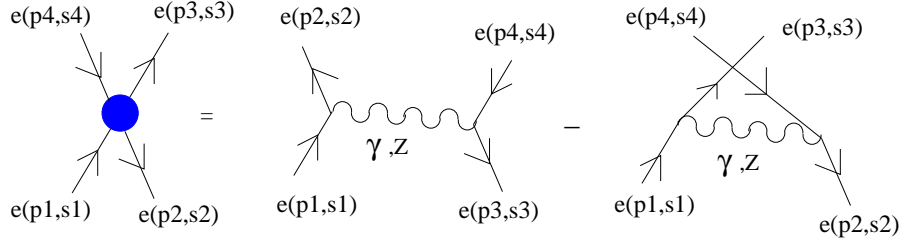}}}
\hspace{3.0cm}
\caption{Feynman diagrams for the Bhabha scattering.}
\protect\label{fig2}
\end{figure}

The scattering amplitude to order $\theta$, for the photon mediated diagram can be written as 
\bea
{\mathcal{A}}_{B}^\gamma = e^2 \left[(1 + \frac{i}{2} p_2 \theta p_1 + \frac{i}{2} p_4 \theta p_3) {\overline v}(p_2,s_2) \gamma^\mu u(p_1,s_1)  {\overline u}(p_3,s_3) \gamma_\mu v(p_4,s_4) \left(\frac{i}{s}\right) \right] \nonumber \\ 
- e^2 \left[ (1 - \frac{i}{2} p_3 \theta p_1 - \frac{i}{2} p_4 \theta p_2) {\overline u}(p_3,s_3) \gamma^\mu u(p_1,s_1)  {\overline v}(p_2,s_2) \gamma_\mu v(p_4,s_4) \left(\frac{i}{t}\right)\right] \nonumber \\
= {\mathcal{A}}_{1B}^\gamma - {\mathcal{A}}_{2B}^\gamma,
\eea
and for the $Z$ mediated diagram as
\bea
{\mathcal{A}}_{B}^Z = \frac{e^2}{x_W^2} \left[(1 + \frac{i}{2} p_2 \theta p_1 + \frac{i}{2} p_4 \theta p_3) {\overline v}(p_2,s_2) \gamma^\mu \Gamma_A^- u(p_1,s_1)  {\overline u}(p_3,s_3) \gamma_\mu \Gamma_A^- v(p_4,s_4) 
\left(\frac{i}{s - m_Z^2}\right) \right] \nonumber \\
- \frac{e^2}{x_W^2} \left[ (1 - \frac{i}{2} p_3 \theta p_1 - \frac{i}{2} p_4 \theta p_2) {\overline u}(p_3,s_3) \gamma^\mu \Gamma_A^- v(p_1,s_1)  {\overline v}(p_2,s_2) \gamma_\mu \Gamma_A^- v(p_4,s_4) \left(\frac{i}{t - m_Z^2}\right)\right] \nonumber \\
= {\mathcal{A}}_{1B}^Z - {\mathcal{A}}_{2B}^Z. \nonumber \\
\eea
The spin-averaged and summed squared-amplitude can be written as 
\bea \label{eqn:BhabhaAmp}
{\overline {|{\mathcal{A}}_{B}|^2}} = {\overline {|{\mathcal{A}}_{1B}^\gamma|^2}} + {\overline {|{\mathcal{A}}_{2B}^\gamma|^2}}+ {\overline {|{\mathcal{A}}_{1B}^Z|^2}} + {\overline {|{\mathcal{A}}_{2B}^Z|^2}} - 2 {\overline {Re({\mathcal{A}}_{1B}^\gamma {\mathcal{A}}_{2B}^{\gamma *})}} - 2 {\overline {Re({\mathcal{A}}_{1B}^Z {\mathcal{A}}_{2B}^{Z *})}} + 2 {\overline {Re({\mathcal{A}}_{1B}^\gamma {\mathcal{A}}_{1B}^{Z *})}} 
\nonumber \\
 - 2 {\overline {Re({\mathcal{A}}_{1B}^\gamma {\mathcal{A}}_{2B}^{Z *})}} - 2 {\overline {Re({\mathcal{A}}_{2B}^\gamma {\mathcal{A}}_{1B}^{Z *})}} + 2 {\overline {Re({\mathcal{A}}_{2B}^\gamma {\mathcal{A}}_{2B}^{Z *})}} = \frac{1}{4} \sum_{spins} |{\mathcal{A}}_B^\gamma + {\mathcal{A}}_B^Z |^2. \nonumber \\
\eea
Several terms in the squared-amplitude are given in Appendix C.

The differential cross-section of a $2 \to 2$  scattering
(M\"{o}ller or Bhabha) reads as
\bea
\frac{d \sigma}{d \Omega} = \frac{1}{64 \pi^2 s^2} \beta ~{\overline {|{\mathcal{A}}|^2}} = \frac{1}{64 \pi^2 s} {\overline {|{\mathcal{A}}|^2}}
\eea
where $\beta = s (1 - 4 m^2/s)^{1/2} \simeq s$(since $\sqrt{s} \gg 2 m$) and $d \Omega = d cos\theta^* d\phi$. The amplitude square ${|{\mathcal{A}}|^2}$ corresponds to ${|{\mathcal{A}}_M|^2}$ for M\"{o}ller scattering and ${|{\mathcal{A}}_B|^2}$ for Bhabha scattering. 
\section{Numerical Analysis}
After obtaining the angular distributions of the diffrential cross-section in the presence of space-time noncommutativity, we then 
analyze the distributions.  In our analysis, we set the machine energy at $\sqrt{s}(=E_{com}) = 1500$ GeV.
\subsection{Angular distribution of the M\"{o}ller scattering in the NCSM}
In Figs. 3(a,b) we have shown the angular distribution 
$\frac{d \sigma}{d \Omega}$ as a function of the azumuthal angle 
$\phi$ with $\theta^*$(polar angle) being fixed at 
$\pi/4$ and $3 \pi/4$, respectively. 
In the usual SM, the azumuthal distribution is supposed to be flat and the lowest horizontal curve in Figs. 3a and 3b, corresponds to that. Note the differences between figures for different $\theta^*$ and among different curves for a given $\theta^*$. In Fig. 3a(or 3b), the topmost curve corresponds to $\Lambda_{NC} = 800$ GeV and this differs maximally from the CSM flat curve. The 2nd and 3rd curves (moving from the top) stands for
$\Lambda_{NC} = 1000$ and $1200$ GeV and they still lie above the flat curve. This is essentially due to the factor 
$cos\phi + sin\phi$ whose origin is in the 
\newpage
\begin{figure}
\subfigure[]{
\label{PictureThreeLabel}
\hspace*{-0.75 in}
\begin{minipage}[b]{0.5\textwidth}
\centering
\includegraphics[width=\textwidth]{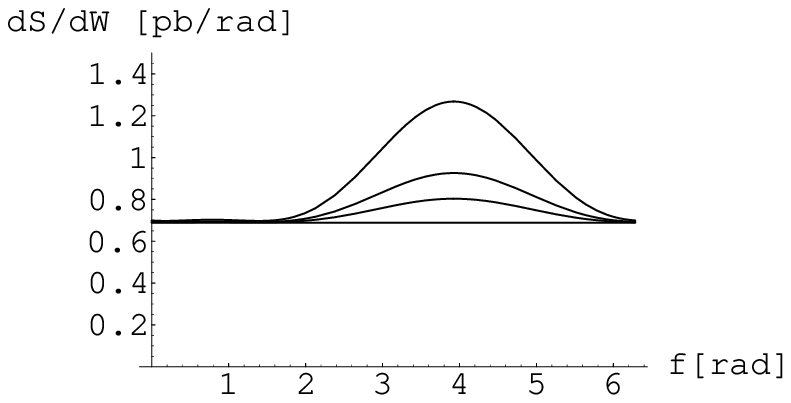}
\end{minipage}}
\subfigure[]{
\label{PictureFourLabel}
\hspace*{0.5in}
\begin{minipage}[b]{0.5\textwidth}
\centering
\includegraphics[width=\textwidth]{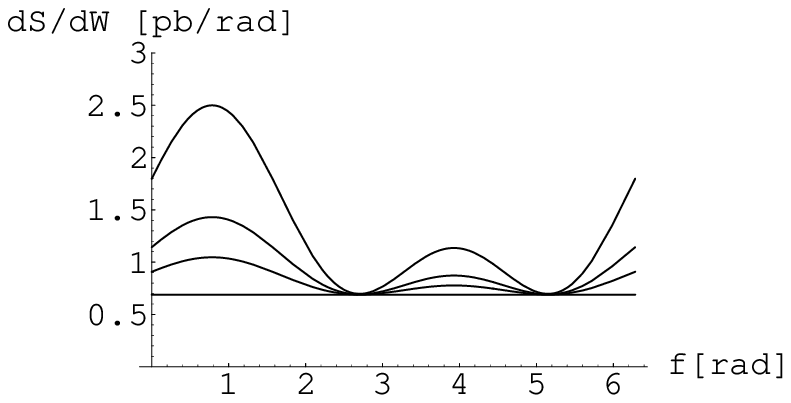}
\end{minipage}}
\end{figure}
\vspace*{-0.5in}
\noindent {\bf FIG. 3.}
{{\it The $\frac{d\sigma}{d\Omega}$ (pb/radian) distribution (for fixed $\theta^*$) as a function of the $\phi$ (in radian) for M\"{o}ller scattering. $\theta^*$ is chosen as $\pi/4$ and $3 \pi/4$, respectively. The c.o.m energy($E_{com}$) is fixed at $\sqrt{s}= 1500$ GeV. The lowest horizontal curve is due to the commutative SM. Note that $\frac{d\sigma}{d\Omega}$ and $\phi$ appear, respectively, as $ds/dW$ and $f$ in these and later figures.}}

\noindent noncommutativity of space-time and which is simply the identity in the CSM. Note that the deviation from the CSM prediction is maximal in a wide region (from small to large values) of $\phi$ (see Figs. 3a and 3b). 
In Figs. 3(c)-3(f), $d\sigma/d\Omega$ is plotted as a function of 
$cos\theta^*$ with $\phi$ being fixed at $=0, \pi/2, \pi$ and $3 \pi/2$, respectively. As before, the lowest curve in each of these figures  is due to the CSM, whereas the topmost, next to the top and next to that (second from the lowest one) respectively stands for $\Lambda_{NC} = 800, 1000$ and $1200$. Differences between these figures are worthwhile to note. 
 Obviously the distribution is asymmetric around $cos\theta^* =0$ (which means $\theta^* = \pi/2$) axis. Such an asymmetry which is non-vanishing in the CSM due to the $\gamma$ and $Z$ amplitude interference terms(the lowest curve corresponds to that), increases with the decrease of $\Lambda_{NC}$. For example the asymmetry corresponding to
$\Lambda_{NC} = 800$ GeV is greater than that obtained at $\Lambda_{NC} = 1200$. So the noncommutative geometry does indeed have an impact on the  $cos\theta^*$ distribution and and thus on the direct CP asymmetry $A_{CP}$ in case of M\"{o}ller scattering. Note that the earlier authors \cite{Hewett01} while studying the M\"{o}ller scattering, did not study such an asymmetry, which we did and is one of our main result in this work. 
 

\newpage
\begin{figure}
\subfigure[]{
\label{PictureThreeLabel}
\hspace*{-0.75 in}
\begin{minipage}[b]{0.5\textwidth}
\centering
\includegraphics[width=\textwidth]{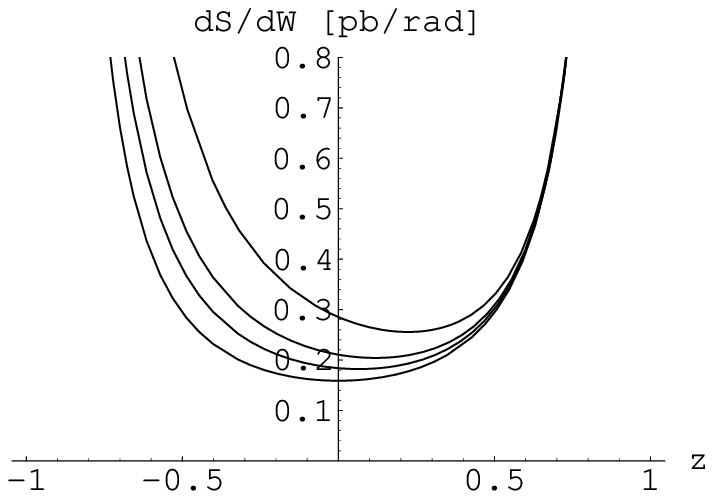}
\end{minipage}}
\subfigure[]{
\label{PictureFourLabel}
\hspace*{0.5in}
\begin{minipage}[b]{0.5\textwidth}
\centering
\includegraphics[width=\textwidth]{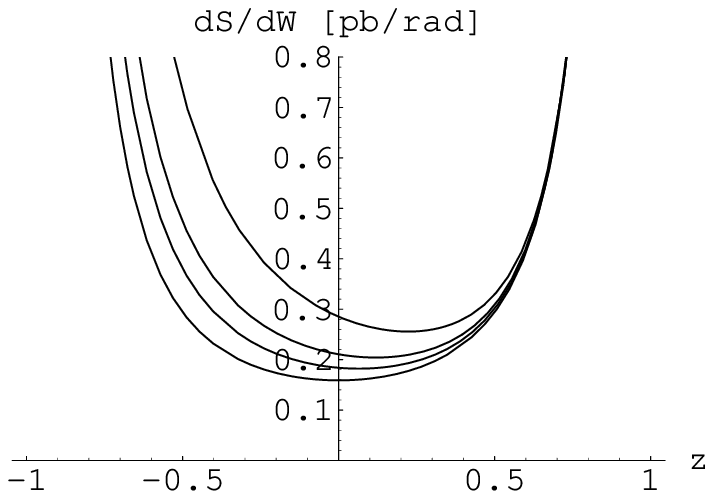}
\end{minipage}}
\subfigure[]{
\label{PictureThreeLabel}
\hspace*{-0.75 in}
\begin{minipage}[b]{0.5\textwidth}
\centering
\includegraphics[width=\textwidth]{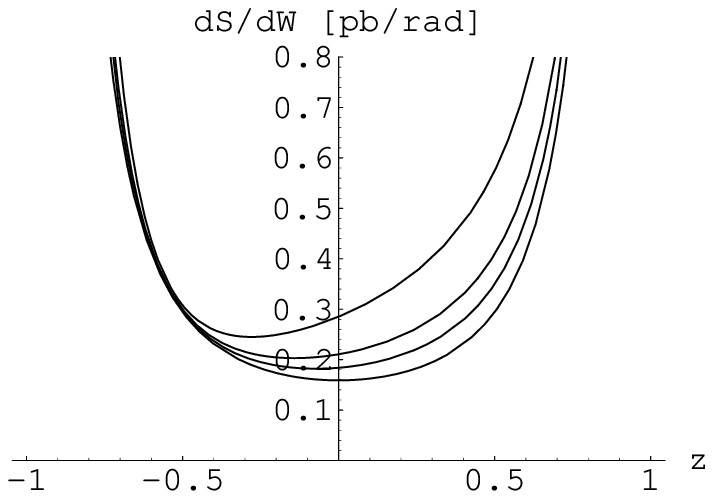}
\end{minipage}}
\subfigure[]{
\label{PictureFourLabel}
\hspace*{0.5in}
\begin{minipage}[b]{0.5\textwidth}
\centering
\includegraphics[width=\textwidth]{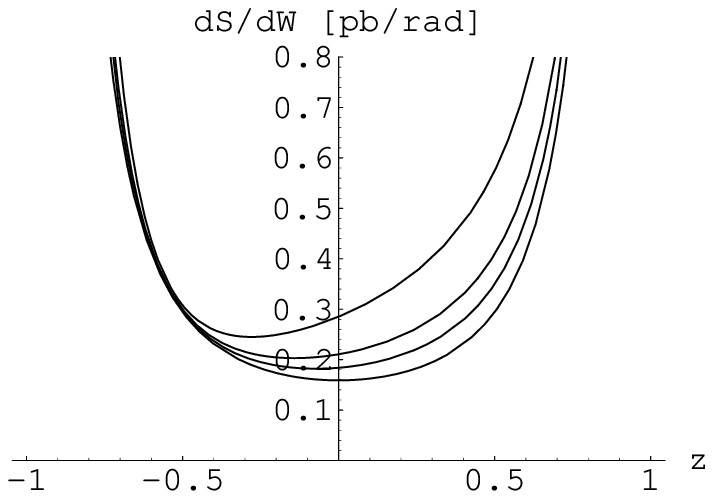}
\end{minipage}}
\end{figure}
\vspace*{-0.5in}
\noindent {\bf FIG. 3.}
{{\it The polar distribution $\frac{d\sigma}{d\Omega}$ (pb/radian) as a function of the $cos\theta^*$(in radian) for the M\"{o}ller scattering. The azumuthal angle $\phi$ is kept fixed at $0, \pi/2, 3 \pi$ and $3 \pi/2$. 
The lowest curve in each of these figures is due to the commutative SM. Note that $cos\theta^*$ appear as $z$ in these and later figures.}}
\subsection{Angular distribution of the Bhabha scattering in the NCSM}
We next turn our attention to the Bhabha scattering and examine the impact of space-time noncommutativity on the angular distribution. 
In Figs. 3g and 3h, we have plotted the distribution $\frac{d \sigma}{d \Omega}$ as a function of the azumuthal angle $\phi$ for fixed $\theta^*$ and it is $\pi/4$ for Fig. 3g and $3 \pi/4$ for Fig. 3h, respectively. The azumuthal distribution in the usual SM(CSM) is completely flat and in each figure the lowest flat curve resembles to that.  Note the differences
\newpage
\begin{figure}
\subfigure[]{
\label{PictureThreeLabel}
\hspace*{-0.75 in}
\begin{minipage}[b]{0.5\textwidth}
\centering
\includegraphics[width=\textwidth]{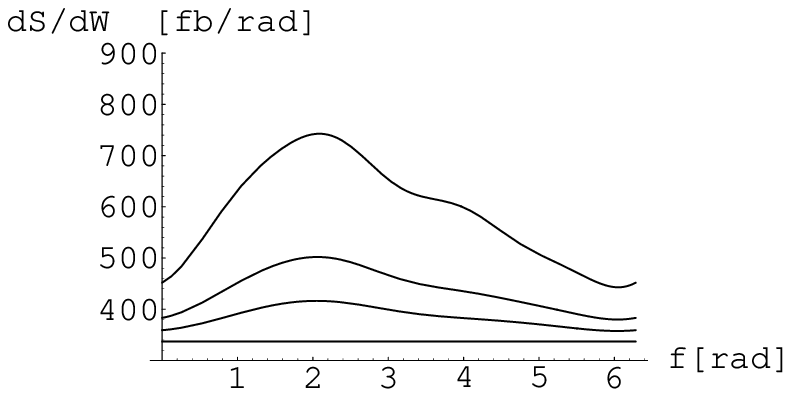}
\end{minipage}}
\subfigure[]{
\label{PictureFourLabel}
\hspace*{0.5in}
\begin{minipage}[b]{0.5\textwidth}
\centering
\includegraphics[width=\textwidth]{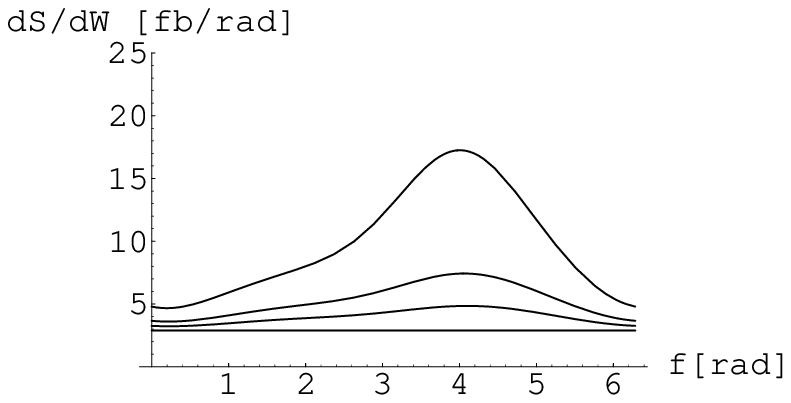}
\end{minipage}}
\end{figure}
\vspace*{-0.5in}
\noindent {\bf FIG. 3.}
{{\it The $\frac{d\sigma}{d\Omega}$ (fb/radian) distribution (for fixed $\theta^*$) as a function of the $\phi$ (in radian) for Bhabha scattering. $\theta^*$ is fixed at $\pi/4$ and $3 \pi/4$. 
We choose the c.o.m energy($E_{com}$) equal to 
$\sqrt{s}= 1500$ GeV. The lowest horizontal curve is the commutative SM result.}}

\noindent between figures for different $\theta^*$ (i.e. Figs. 3g and 3h) and among different curves for a given $\theta^*$ (i.e. in Fig. 3g or 3h). 
In Fig. 3g(or 3h), the topmost curve corresponds to $\Lambda_{NC} = 800$ GeV and the 2nd and 3rd one (starting from the topmost one) stands for
$\Lambda_{NC} = 1000$ GeV and $1200$ GeV, respectively. 
 Note that the topmost curve(in either Fig.) differs maximally from the CSM flat curve and the other two still lie above the 
flat curve and this is due to the same $\phi$ dependent factor as discussed in earlier section. Also note that the deviation from the CSM flat curve is maximal in a wide region (from small to large values) of $\phi$ (see Figs. 3g and 3h).   

In Figs. 3(i)-3(l) we have plotted $d\sigma/d\Omega$ as a function of 
$cos\theta^*$ for fixed $\phi$. $\phi$ is chosen as $=0, \pi/2, \pi$ and $3 \pi/2$, respectively. The lowest 
curve in each of these figures  is due to the CSM, whereas the topmost, next to the top and next to that (second from the lowest one) respectively stands for $\Lambda_{NC} = 800, 1000$ and $1200$. Obviously the distribution which is asymmetric around $cos\theta^* =0$ (i.e. $\theta^* = \pi/2$) axis, as is expected within the CSM due to the $\gamma$ and $Z$ interference terms, increases with the decrease of $\Lambda_{NC}$. The lowest curve(in each of these figures) corresponds to that due to the CSM. Note the change in asymmetry prediction with the change in fixed 
$\phi$ value which is minimum for $\phi=0$ and maximum for $\phi=\pi/2$. 
Also to note that the asymmetry at 
$\Lambda_{NC} = 800$ GeV is greater than that obtained at $\Lambda_{NC} = 1200$ in a given curve. So the noncommutative geometry does indeed have an 
 
\newpage
\begin{figure}
\subfigure[]{
\label{PictureThreeLabel}
\hspace*{-0.75 in}
\begin{minipage}[b]{0.5\textwidth}
\centering
\includegraphics[width=\textwidth]{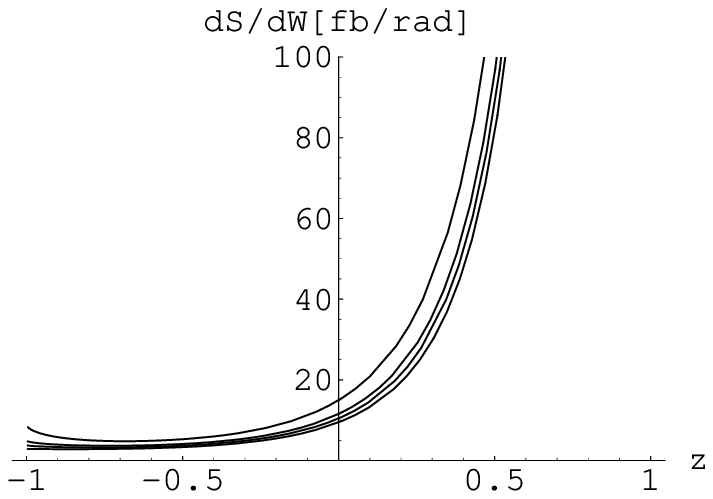}
\end{minipage}}
\subfigure[]{
\label{PictureFourLabel}
\hspace*{0.5in}
\begin{minipage}[b]{0.5\textwidth}
\centering
\includegraphics[width=\textwidth]{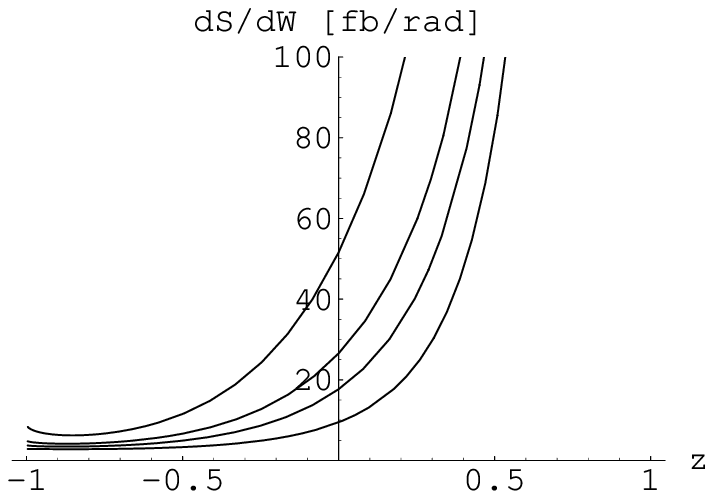}
\end{minipage}}
\subfigure[]{
\label{PictureThreeLabel}
\hspace*{-0.75 in}
\begin{minipage}[b]{0.5\textwidth}
\centering
\includegraphics[width=\textwidth]{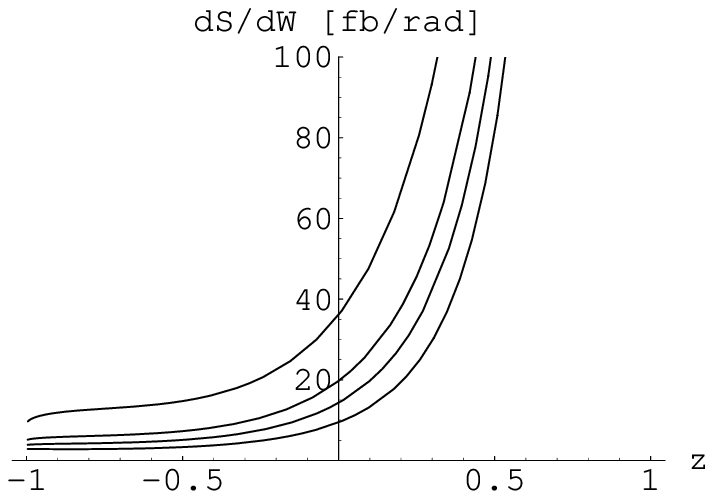}
\end{minipage}}
\subfigure[]{
\label{PictureFourLabel}
\hspace*{0.5in}
\begin{minipage}[b]{0.5\textwidth}
\centering
\includegraphics[width=\textwidth]{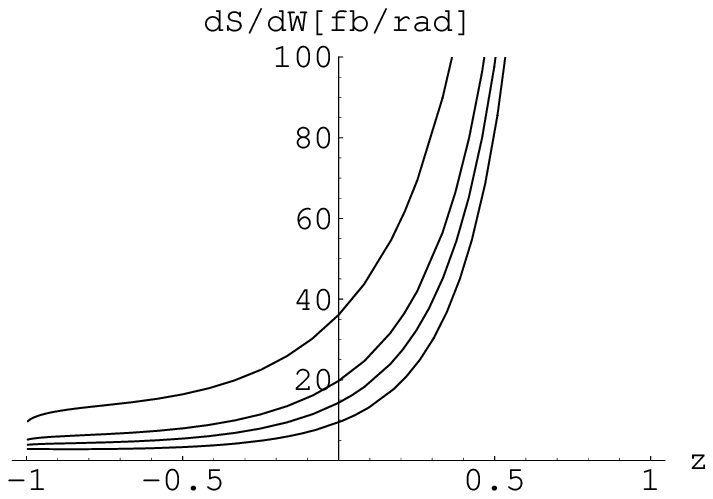}
\end{minipage}}
\end{figure}
\vspace*{-0.5in}
\noindent {\bf FIG. 3.}
{{\it The polar distribution $\frac{d\sigma}{d\Omega}$ (fb/radian) as a function of the $cos\theta^*$ (in radian) for the Bhabha scattering. The azumuthal angle $\phi$ is kept fixed at $0, \pi/2, 3 \pi$ and $3 \pi/2$, respectively. The lowest curve in each of these figures is due to the commutative SM.}}

\noindent impact on the  $cos\theta^*$ distribution and and thus on $A_{CP}$ in case of Bhabha scattering. Note that the earlier authors \cite{Hewett01} did not include $Z$-mediated diagram while studying the Bhabha scattering and thus didn't pay attention on such asymmetry plot.
We consider the $Z$ mediated amplitude and the impact of the space-time noncommutativity on such an asymmetry plot and is one of the main result of the present  work. 
\section{Conclusion}
The idea that around the TeV scale the space and time coordinates no longer remains commutative in nature (and becomes non-commutative) is explored here by investigating its impact on the two fundamental processes, the M\"{o}ller and Bhabha scattering. In addition to the photon exchange, we considered here also 
the $Z$ boson exchange, which leads to substantial amount of modification of  earlier works(in which $Z$-exchange was not considered). From the azumuthal distribution $d\sigma/d\Omega$ for a fixed $\theta^*$ we see that for certain $\theta^*$ ($\pi/4$ and $3 \pi/4$) the distribution differs significantly from the one expected from the commutative version of the SM. At around $\Lambda_{NC} \sim 1500$ (noting that $E_{cm} = 1500$~GeV in this analysis) the NC effect smeared out, what remains is the one(the flat curve) 
as expected within the CSM. With the lowering of $\Lambda_{NC}$, say for example from $1200$ GeV to $500$ GeV, the NC effect gets enhanced. Besides the azumuthal distribution, we also obtain the polar distribution $d\sigma/d\Omega$ for a fixed $\phi$ and hence the direct CP-asymmetry. We observed that such distribution  do really depend on $\theta^*$ and have 
an asymmetry which simply got an enhancement for certain $\phi$ value (e.g. $\phi = \pi/2$ and $3 \pi/2$). Such a result is completely new.   
Combining the azumuthal and polar distribution, we found that for the c.o.m energy $E_{cm}= 1500$~GeV, the noncommutative effect do manifests itself at around $\Lambda_{NC} \sim 800$ to $1200$ GeV and differs significantly from 
the curve (the lowest horizontal curve in each plot) obtained from the commutative version of the SM.   
\begin{acknowledgments}
P.~K.~Das would like to thank Prof T.R.Govindrajan of IMSc,Chennai for useful discussions during his stay at IMSc. The work of N.G.Deshpande is supported by 
the US DOE under Grant No. DE-FG02-96ER40969.
\end{acknowledgments}
\appendix

\section{Feynman rules to order ${\mathcal{O}}(\theta)$ }
We follow the Reference \cite{Melic:2005ep} for the Feynman rule for several interactions, propagators.
The Feynman rule for the $e (p_{in})-e(p_{out}) - \gamma(k)$ vertex to 
$O(\theta)$ is 
\bea
i e Q_f \left[\gamma_\mu - \frac{i}{2} k^\nu (\theta_{\mu \nu \rho} p^\rho_{in} - \theta_{\mu \nu} m_f ) \right]
\nonumber \\
= i e Q_f \gamma_\mu + \frac{1}{2} e Q_f \left[(p_{out} \theta p_{in}) \gamma_\mu - (p_{put} \theta)_\mu (\pinsla -  m_f) - (\poutsla -  m_f) (\theta p_{in})_\mu \right], \nonumber 
\eea
and for the vertex $e(p_{in})-e(p_{out}) - Z(k)$ is 
\bea
\frac{e}{sin2\theta_W} \left[i \gamma_\mu \Gamma_A^- \right] 
+ \frac{e}{2 sin2\theta_W} \left[(p_{out} \theta p_{in}) \gamma_\mu \Gamma_A^- 
 - (p_{put} \theta)_\mu \Gamma_A^+ (\pinsla -  m_f) - (\poutsla -  m_f) 
\Gamma_A^- (\theta p_{in})_\mu \right] \nonumber .
\eea
Here $\theta_{\mu \nu \rho}= \theta_{\mu \nu} \gamma_\rho + \theta_{\nu \rho}
\gamma_\mu + \theta_{\rho \mu} \gamma_\nu$,~ 
$\Gamma_A^\pm = (c_V^e \pm c_A^e \gamma_5)$ and  $Q_f = \mp 1$ for $e^{\mp}$. \\
 Also $ p_{out} \theta p_{in} = p_{out}^\mu  \theta_{\mu \nu}  p_{in}^\nu = -p_{in} \theta p_{out}$. The momentum conservation reads as 
$p_{in} + k = p_{out}$. 

\section{Squared amplitude of the M\"{o}ller ($e^-(p_1) e^-(p_2) \to e^-(p_3) e^-(p_4)$ scattering}

Here we explicitly present several squared-amplitude terms of Eq.(\ref{eqn:MollerAmp}). Defining $p_1$, $p_2$, $p_3$ and $p_4$ to be the momenta 
of the of the initial and final state electrons, the terms in the squared matrix element are given by 
\bea
{\overline {|{\mathcal{A}}_{1M}^\gamma|^2}} &=& \frac{e^4}{4 t^2} 
(1 + \frac{1}{4}A^2)
Tr[\ponesla \gamma_\nu \pthreesla \gamma_\mu]  \times Tr[\ptwosla \gamma^\nu \pfoursla \gamma^\mu], \nonumber \\
{\overline {|{\mathcal{A}}_{2M}^\gamma|^2}} &=& \frac{e^4}{4 u^2} 
(1 + \frac{1}{4}B^2) 
Tr[\ponesla \gamma_\nu \pthreesla \gamma_\mu] \times 
Tr[\ptwosla \gamma^\nu \pfoursla \gamma^\mu], 
\nonumber \\
{\overline {|{\mathcal{A}}_{1M}^Z|^2}} &=& \frac{e^4}{4 x_W^4 (t -m_Z^2)^2} 
(1 + \frac{1}{4}A^2) 
Tr[\ponesla \gamma_\nu \Gamma_A^- \pthreesla \gamma_\mu \Gamma_A^-] \times 
Tr[\ptwosla \gamma^\nu \Gamma_A^- \pfoursla \gamma^\mu \Gamma_A^-], \nonumber \\
{\overline {|{\mathcal{A}}_{2M}^Z|^2}} &=& \frac{e^4}{4 x_W^4 (u -m_Z^2)^2} 
(1 + \frac{1}{4}B^2)
Tr[\ponesla \gamma_\nu \Gamma_A^- \pfoursla \gamma_\mu \Gamma_A^-] \times 
Tr[\ptwosla \gamma^\nu \Gamma_A^- \pthreesla \gamma^\mu \Gamma_A^-], \nonumber \\
- 2 {\overline {Re({\mathcal{A}}_{1M}^\gamma {\mathcal{A}}_{2M}^{\gamma *})}} 
&=&  - \frac{e^4}{2 u t} Re \left[(1 + \frac{i}{2}A) (1 - \frac{i}{2}B)  Tr[\ponesla \gamma_\nu \pfoursla \gamma_\mu \ptwosla \gamma^\nu \pthreesla \gamma^\mu]\right], \nonumber \\
- 2 {\overline {Re({\mathcal{A}}_{1M}^Z {\mathcal{A}}_{2M}^{Z *})}}
&=&  - \frac{e^4}{2 x_W^4 (u-m_Z^2) (t-m_Z^2)} Re \left[ 
(1 + \frac{i}{2}A) (1 - \frac{i}{2}B) 
Tr[\ponesla \gamma_\nu \Gamma_A^- \pfoursla \gamma_\mu \Gamma_A^- \ptwosla \gamma^\nu \Gamma_A^- \pthreesla \gamma^\mu \Gamma_A^-] \right], \nonumber \\
+ 2 {\overline {Re({\mathcal{A}}_{1M}^\gamma {\mathcal{A}}_{1M}^{Z *})}}
&=&   \frac{e^4}{2 x_W^2 t (t-m_Z^2)} Re \left[(1 + \frac{1}{4}A^2)
Tr[\ponesla \gamma_\nu \Gamma_A^- \pthreesla \gamma_\mu]
Tr[\ptwosla \gamma^\nu \Gamma_A^- \pfoursla \gamma^\mu] \right], \nonumber \\
- 2 {\overline {Re({\mathcal{A}}_{1M}^\gamma {\mathcal{A}}_{2M}^{Z *})}}
&=&   -\frac{e^4}{2 x_W^2 t (u-m_Z^2)} Re \left[(1 + \frac{i}{2}A) 
(1 - \frac{i}{2}B) 
Tr[\ponesla \gamma_\nu \Gamma_A^- \pfoursla \gamma_\mu \ptwosla \gamma^\nu 
\Gamma_A^- \pthreesla \gamma^\mu] \right], \nonumber \\
- 2 {\overline {Re({\mathcal{A}}_{2M}^\gamma {\mathcal{A}}_{1M}^{Z *})}}
&=&   -\frac{e^4}{2 x_W^2 u (t-m_Z^2)} Re \left[(1 + \frac{i}{2}A) 
(1 - \frac{i}{2}B) 
Tr[\ponesla \gamma_\nu \Gamma_A^- \pthreesla \gamma_\mu \ptwosla \gamma^\nu 
\Gamma_A^- \pfoursla \gamma^\mu] \right], \nonumber \\
+ 2 {\overline {Re({\mathcal{A}}_{2M}^\gamma {\mathcal{A}}_{2M}^{Z *})}}
&=&   \frac{e^4}{2 x_W^2 u (u-m_Z^2)} Re \left[(1 + \frac{1}{4}B^2)
Tr[\ponesla \gamma_\nu \Gamma_A^- \pfoursla \gamma_\mu]
Tr[\ptwosla \gamma^\nu \Gamma_A^- \pthreesla \gamma^\mu] \right], \nonumber \\
\eea
where $ A=(p_3 \theta p_1 - p_4 \theta p_2)$ and 
$ B= (p_4 \theta p_1 - p_3 \theta p_2).$ \\
In the c.o.m frame of $e^-(p_1)$ and $e^-(p_2)$ collision, the following prescription for the $4$-momenta are used:
\bea
p_1 &=& \left(\frac{\sqrt{s}}{2}, 0, 0, \frac{\sqrt{s}}{2}\right)=(E_1, \vec{P_1}), \nonumber \\
p_2 &=& \left(\frac{\sqrt{s}}{2}, 0, 0, -\frac{\sqrt{s}}{2}\right)=(E_2, \vec{P_2}), \nonumber \\
p_3&=& \left(\frac{\sqrt{s}}{2},\frac{\sqrt{s}}{2} sin\theta^* cos\phi,\frac{\sqrt{s}}{2} sin\theta^* sin\phi, \frac{\sqrt{s}}{2} cos\theta^* \right)  = (E_3, \vec{P_3}), \nonumber \\
p_4&=& \left(\frac{\sqrt{s}}{2},-\frac{\sqrt{s}}{2} sin\theta^* cos\phi,-\frac{\sqrt{s}}{2} sin\theta^* sin\phi, -\frac{\sqrt{s}}{2} cos\theta^* \right) = (E_4, \vec{P_4}),
\eea
where $\theta^*$ is the scattering angle made by the $3$-momentum vector $p_3$ of $e^-(p_3)$ with the z axis and $\phi$ is the azumuthal angle. In above 
$m = m_e \simeq 0$.
Note that $\vec{P_1} = |P_1| \hat{z}$, $\vec{P_2} = -|P_2| \hat{z}$ and
$\vec{P_1} + \vec{P_2} = 0 = \vec{P_3} + \vec{P_4}$. In the relativistic limit($s \gg 4 m^2$), we find $s=(E_1 + E_2)^2 = 4 E^2$ (with $E=E_1=E_2$),
$t = -\frac{s}{2} (1 - cos\theta^*)$ and $u = -\frac{s}{2} (1 + cos\theta^*)$.

Writing $\theta_{\mu\nu}$ as $c_{\mu \nu}/{\Lambda_{NC}^2}$ and taking all the nonvanishing $c_{\mu\nu}$ to be unity \cite{Hewett01},
we evaluate the quantities appearing in the squared-amplitude as
\bea
p_3 \theta p_1 - p_4 \theta p_2 = \frac{s}{2 \Lambda_{NC}^2} \left[1 - cos \theta^* - sin \theta^* (cos \phi + sin \phi) \right], \nonumber \\
p_4 \theta p_1 - p_3 \theta p_2 = \frac{s}{2 \Lambda_{NC}^2} \left[1 + cos \theta^* + sin \theta^* (cos \phi + sin \phi) \right].
\eea

\section{Squared amplitude of the Bhabha ($e^-(p_1) e^+(p_2) \to e^-(p_3) e^+(p_4)$) scattering }
Here we present several squared-amplitude terms of Eq.(\ref{eqn:BhabhaAmp}). 
We define $p_1$, $p_2$ and $p_3$, $p_4$ are the $4$-momenta of the initial 
$e^-$ and $e^+$ and the final $e^-$ and $e^+$. The terms in the squared matrix element are given by
\bea
{\overline {|{\mathcal{A}}_{1B}^\gamma|^2}} &=& \frac{e^4}{4 s^2} 
(1 + \frac{1}{4}E^2) 
Tr[\ptwosla \gamma_\mu \ponesla \gamma_\nu]  \times Tr[\pthreesla \gamma^\mu \pfoursla \gamma^\nu], \nonumber \\
{\overline {|{\mathcal{A}}_{2B}^\gamma|^2}} &=& \frac{e^4}{4 t^2} (1 + \frac{1}{4}F^2)
Tr[\ponesla \gamma_\nu \pthreesla \gamma_\mu] \times 
Tr[\pfoursla \gamma^\nu \ptwosla \gamma^\mu], 
\nonumber \\
{\overline {|{\mathcal{A}}_{1B}^Z|^2}} &=& \frac{e^4}{4 x_W^4 (s -m_Z^2)^2} 
(1 + \frac{1}{4}E^2) 
Tr[\ponesla \gamma_\nu \Gamma_A^- \ptwosla \gamma_\mu \Gamma_A^-] \times 
Tr[\pfoursla \gamma^\nu \Gamma_A^- \pthreesla \gamma^\mu \Gamma_A^-], \nonumber \\
{\overline {|{\mathcal{A}}_{2B}^Z|^2}} &=& \frac{e^4}{4 x_W^4 (t -m_Z^2)^2})
(1 + \frac{1}{4}F^2)
Tr[\ponesla \gamma_\nu \Gamma_A^- \pthreesla \gamma_\mu \Gamma_A^-] \times 
Tr[\pfoursla \gamma^\nu \Gamma_A^- \ptwosla \gamma^\mu \Gamma_A^-], \nonumber \\
- 2 {\overline {Re({\mathcal{A}}_{1B}^\gamma {\mathcal{A}}_{2B}^{\gamma *})}} 
&=&  - \frac{e^4}{2 s t} Re \left[(1 + \frac{i}{2}E) (1 + \frac{i}{2}F)  Tr[\ponesla \gamma_\nu \pthreesla 
\gamma^\mu \pfoursla \gamma^\nu \ptwosla \gamma_\mu]\right], \nonumber \\
+ 2 {\overline {Re({\mathcal{A}}_{1B}^\gamma {\mathcal{A}}_{1B}^{Z *})}}
&=&   \frac{e^4}{2 x_W^2 s (t-m_Z^2)} Re \left[(1 + \frac{1}{4}E^2)
Tr[\ponesla \gamma_\nu \Gamma_A^- \ptwosla \gamma_\mu]
Tr[\pfoursla \gamma^\nu \Gamma_A^- \pthreesla \gamma^\mu] \right], \nonumber \\
- 2 {\overline {Re({\mathcal{A}}_{1B}^\gamma {\mathcal{A}}_{2B}^{Z *})}}
&=&   -\frac{e^4}{2 x_W^2 s (t-m_Z^2)} Re \left[ (1 + \frac{i}{2}E) (1 + \frac{i}{2}F)
Tr[\ponesla \gamma_\nu \Gamma_A^- \pthreesla \gamma_\mu \pfoursla \gamma^\nu 
\Gamma_A^- \ptwosla \gamma^\mu] \right], \nonumber \\
- 2 {\overline {Re({\mathcal{A}}_{2B}^\gamma {\mathcal{A}}_{1B}^{Z *})}}
&=&   -\frac{e^4}{2 x_W^2 t (s-m_Z^2)} Re \left[ (1 - \frac{i}{2}E) 
(1 - \frac{i}{2}F)
Tr[\ponesla \gamma_\nu \Gamma_A^- \ptwosla \gamma^\mu \pfoursla \gamma^\nu 
\Gamma_A^- \pthreesla \gamma_\mu] \right], \nonumber \\
+ 2 {\overline {Re({\mathcal{A}}_{2B}^\gamma {\mathcal{A}}_{2B}^{Z *})}}
&=&   \frac{e^4}{2 x_W^2 u (u-m_Z^2)} Re \left[(1 + \frac{1}{4}F^2)
Tr[\ponesla \gamma_\nu \Gamma_A^- \pfoursla \gamma_\mu]
Tr[\ptwosla \gamma^\nu \Gamma_A^- \pthreesla \gamma^\mu] \right], \nonumber \\
- 2 {\overline {Re({\mathcal{A}}_{1B}^Z {\mathcal{A}}_{2B}^{Z *})}}
&=&  - \frac{e^4}{2 x_W^4 (u-m_Z^2) (t-m_Z^2)} Re \left[ (1 + \frac{i}{2}E) 
(1 + \frac{i}{2}F)
Tr[\ponesla \gamma_\nu \Gamma_A^- \pfoursla \gamma_\mu \Gamma_A^- \ptwosla \gamma^\nu \Gamma_A^- \pthreesla \gamma_\mu \Gamma_A^-] \right], \nonumber \\
\eea
where $ E=(p_2 \theta p_1 + p_4 \theta p_3),~
F=(p_3 \theta p_1 + p_4 \theta p_2). \nonumber$
In the c.o.m frame of $e^-(p_1) - e^+(p_2)$ collision, with the $4$-momenta prescription as discussed above, we then evaluate the quantities
\bea
p_2 \theta p_1 + p_4 \theta p_3 &=& \frac{s}{2 \Lambda_{NC}^2} \left[1 + cos \theta^* + sin \theta^* (cos \phi + sin \phi) \right], \nonumber \\
p_3 \theta p_1 + p_4 \theta p_2 &=& \frac{s}{2 \Lambda_{NC}^2} \left[ sin \theta^* (cos \phi + sin \phi) \right].
\eea


\newpage
\begin{thebibliography}{10}
\bibitem{Snyder47}
H.S. Snyder, \PRD(71,38,1947). 

\bibitem{Connes98} A.Connes,M.R.Douglas and A.Schwarz, \JHEP(02,003,1998).

\bibitem{Douglas98}M.R.Douglas and C.Hull, \JHEP(02,008,1998). 

\bibitem{SW99}
N. Seiberg and E. Witten, \JHEP(09,032,1999). 

\bibitem{Schomerus} V. Schomerus, \JHEP(06,030,1999). 

\bibitem{Jurco}
B.~Jurco, P.~ Schupp and J.~Wess,
\MPL(A16,343,2001).

\bibitem{Witten96}E.Witten, \NP(B471,135,1996);
P.Horava and E.Witten, \NP(B460,506,1996). 

\bibitem{Douglas}M.~R.~Douglas and N.~Nekrasov, \RMP(73,977,2001);~I.~F.~Riad and M.~M.~Sheikh-Jabbari, 
\JHEP(08,45,2000); 
B.~ Jurco and P.~Schupp,
\EPJC(14,367,2000).


\bibitem{Calmet} X.~Calmet, B.~Jurco, P.~Schupp,~J. Wess and M. ~Wohlgenannt,
\EPJC(23,363,2002); 
X.~ Calmet
\EPJC(50,113,2007); X.~Calmet and M.~Wohlgenannt
\PRD(68,025016,2003). 

\bibitem{Hewett01}J.Hewett, F.J.Petriello and T.G.Rizzo, hep-ph/0201275;  J.Hewett \etal, \PRD(64,075012,2001), \PRD(66,036001,2002); T.~G.~Rizzo \IJMP(A18,2797,2003); I.~Hinchliffe, N.~Kersting and Y.~L.~Ma,
\IJMP(A19,179,2004); 
A.~Alboteanu, T.~Ohl and R.~Ruckl,
 \APP(38,3647,2007). 

\bibitem{Melic:2005ep}
  B.~Meli\'{c}, K.~ P.~Kumericki, J.~ Trampetic, P.~ Schupp 
and M.~ Wohlgenannt
\EPJC(42,483,2005);
B.~Meli\'{c} {\it et al.},
\EPJC(42,499,2005). 
P.~Schupp and J.~Trampetic,
Springer Proc.Phys. {\bf 98}, 219 (2005);
P.~Schupp,
Lect. Notes Phys.~{\bf 616}, 305 (2003). 

\bibitem{Trampetic}
G.~Duplancic, P.~Schupp and J.~ Trampetic
\EPJC(32,141,2003); 
W.~Behr, N.~G.~ Deshpande, G.~ Duplancic, P.~ Schupp, 
J.~ Trampetic and J.~ Wess,
\EPJC(29,441,2003); 
P.~Aschieri, B.~Jurco, P.~Schupp and J.~Wess
\NP(B651,45,2003). 
\end{thebibliography}
\end{document}